\documentclass{llncs}

\usepackage{amsmath,amssymb,amsfonts}
\usepackage{graphicx}
\usepackage{times,mathptm}
\usepackage{url}
\usepackage{verbatim}

\begin{document}
\title{Almost Continuous Transformations of Software and Higher-order Dataflow Programming}

\author{Michael Bukatin\inst{1}\and Steve Matthews\inst{2}}

\institute{Nokia Corporation\\
Burlington, Massachusetts, USA\\ 
\email{bukatin@cs.brandeis.edu}
\and
Department of Computer Science\\
University of Warwick\\
Coventry, UK\\
\email{Steve.Matthews@warwick.ac.uk}}

\maketitle

\begin{abstract}
We consider two classes of stream-based computations which admit taking linear combinations of execution runs:
probabilistic sampling and generalized animation. The dataflow architecture is a natural platform for programming
with streams. The presence of linear combinations allows us to introduce the notion of almost continuous transformation
of dataflow graphs. We introduce a new approach to higher-order dataflow programming: a dynamic
dataflow program is a stream of dataflow graphs evolving by almost continuous transformations.
A dynamic dataflow program would typically run while it evolves. We introduce
Fluid, an experimental open  source system for programming with dataflow graphs and almost continuous
transformations.

\end{abstract}

\section{Introduction}

Standard software architectures tend to be too brittle, and  typical software systems are of discrete nature and not sufficiently robust with respect to small changes.

The ability to take linear combinations of execution runs allows one to change software in a continuous fashion. In this paper, we consider two classes
of computations  which admit taking linear combinations of execution runs: probabilistic sampling and generalized animation. Because these two classes
are both stream-based, dataflow programming is a natural programming paradigm for this situation.

Dataflow paradigm develops since at least early 1960s~\cite{JKellyLochbaumVyssotsky}. Many dozens of text-based and visual programming
dataflow systems have been created since then. For a very extensive and still incomplete overview of state of the field ten years ago see~\cite{WJohnstonHannaMillar}.
Many dataflow programming systems are in active use today, and new dataflow programming systems keep emerging.

The descriptions of dataflow formalisms and syntax of visual dataflow programs are sometimes centered around transformations of data streams. 
During our experiments with prototype dataflow systems in this paper we found this practice to be inconvenient. We explored two alternatives.

For the first-order dataflow programming we found it convenient to consider a formalism based on bipartite dataflow graphs with two kinds of
vertices:  nodes representing data streams and nodes representing transformations. Under this approach, the two kinds of nodes are equally
important, and there is often a duality between them.

For the higher-order dataflow programming, we found it convenient to be able to copy subgraphs of a dataflow graph in meaningful ways, and
for that it is convenient to take an approach centered around target nodes. So the datastreams take the central role under this approach, 
and transformations are made subordinate to the streams they produce.

What we have described so far is applicable to any dataflow programming situations. The ability to take linear combinations comes into play
because it enables the following sequence of reversible transformations. We start with the following two steps we call {\em benign discontinuities}. First,
we take a datastream node $A$ and replace it with nodes $B$ and $C$ connected by an identity transformation. Node $B$ inherits from node $A$ its
incoming links and their transformation. Node $C$ inherits from node $A$ its outgoing links. Second, we replace $C=Id(B)$, with
$C= (1-\alpha)\cdot B + \alpha \cdot D$ where $\alpha = 0$, so a zero weight link from some node $D$ to $C$ is added. Finally, we can continuously
vary $\alpha$. 

We call this process of continuously varying linear coefficients in a dataflow graph punctuated by benign discontinuities at a discrete set of moments in time 
an {\em almost continuous transformation}.

The goal of this paper is to give a detailed informal overview of this approach to dataflow graphs and to describe Fluid~\cite{Fluid}, an experimental open source system
providing first-order and higher-order prototype implementations. A system like Fluid should be applicable both to streams of probabilistic samples and
streams of generalized images. The current prototype works with streams of ordinary images.

\subsection{Motivation and Paper Structure}

It became apparent in our earlier work~\cite{MBukatinMatthews} that having an ability to consider linear combinations of execution runs
should be conducive for program learning in the context of genetic programming and higher-order probabilistic sampling. 
It also became apparent to us that two major classes of computations with the ability to take linear combinations of execution runs,
probabilistic sampling and generalized animations, are both stream-based, and hence it is natural to use dataflow architecture
in this context.

Moreover, dataflow architecture is convenient in the context of program learning, because syntax of dataflow diagrams tends to be
tied closer to their semantics than the syntax of programs in more conventional software architectures.

This paper contributes the ability to almost continuously evolve dataflow programs while they are running. This makes it possible
to sample almost continuous trajectories in the space of dataflow programs, in addition to the usual practice of sampling
the program syntax trees, thus enabling us to try new evolutionary and
probabilistic schemas for
program learning.

The paper is structured as follows. Section~\ref{linear} introduces both classes of stream-based computations admitting linear combinations
of execution runs. This section also briefly describes context:  the ability to take linear combinations should be quite useful, if one wants to enrich
genetic programming with mechanisms inspired by regulation of levels of protein expression; there are recent interesting advances in
probabilistic programming within the higher-order ``sampling the samplers" paradigm; and there are deep connections between
the ability to have negative coefficients in
linear models of computations and various mathematical motives which together constitute the so-called ``partial inconsistency landscape"~\cite{MBukatinMatthews}.

Section~\ref{first} describes using bipartite graphs for the first-order dataflow programming and Fluid implementation of this architecture.

Section~\ref{continuous} describes the notion of almost continuous transformations of dataflow graphs and the resulting streams
of dataflow graphs. The Fluid implementation of this architecture is discussed. An important feature of this implementation is that the dataflow program
in question is normally running as it changes in almost continuous fashion.

We find ourselves in a situation where in addition to ordinary datastreams one has datastreams of dataflow graphs available.
This opens a variety of possibilities for higher-order dataflow programming, some of which are discussed in Section~\ref{higher}.

\section{Linear Models of Computations and Evolutionary Programming}\label{linear}

In this Section we are briefly reviewing the relevant material from~\cite{MBukatinMatthews}.

\subsection{Convex Linear Combinations}

Here we are considering linear combinations $(1-\alpha) \cdot A + \alpha \cdot B$, where $0 \leq \alpha \leq 1$.

A generalized image is a set of points, together with any secondary structure on this set, and with points taking
real values. An ordinary monochrome image corresponds to the situation when this secondary structure is
a continuous or discrete rectangle, and for the color image one typically considers the Cartesian product of this
rectangle and the three-element set representing colors, $\{\mathbf{R}, \mathbf{G}, \mathbf{B}\}$.

If two generalized images have the same secondary structure on their points, one can take convex linear
combinations of those images point-wise.

A generalized animation is a function from discrete or continuous time to generalized images with the same secondary structure.

If two generalized animations have the same secondary structure for their images and are synchronized time-wise,
one can take convex linear combinations of those animations by taking convex linear combinations of
their images corresponding to the same point on the time axis.

If two probabilistic samplers produce points independently sampled from distributions $P$ and $Q$ at the rate
of one sample per clock tick, one can obtain a sampler from the convex linear combination of these distributions
by taking the latest sample from $P$ with probability $(1- \alpha)$ and the latest sample from $Q$ otherwise at each clock tick.

\subsection{Evolutionary Programming and Regulation of Levels of Gene Expression}

To quote from~\cite{MBukatinMatthews}: "Biological systems tend to be much more flexible and adaptive with respect to variation. In particular, biological cells are capable of functioning
at wide ranges of the level of expressions of various proteins, which are machines working in parallel. Regulation of the level of expression
of specific proteins is a key element of flexibility of biological systems. It is argued in evolutionary developmental biology
that the flexible architecture together with conservation of core mechanisms is crucial for the observed rate of 
biological evolution~\cite{JGerhartKirschner,MKirschnerGerhart}.
It is suggested that morphology evolves largely by altering the expression of functionally conserved proteins~\cite{SCarrol}."

Computational architectures which admit the notion of linear combination of execution runs with non-negative coefficients are particularly attractive in this sense. 
Then one can regulate the system simply by controlling coefficients in a linear combination of its components,  as
computational equivalent of the level of expression of a particular protein. 

\subsection{Sampling the samplers}\label{sampling_the_samplers}

The term ``higher-order probabilistic programming" usually means a higher-order functional
programming language implementing sampling semantics. We have recently seen examples of research
implementing higher-order sampling schemas in a more narrow and focused sense of the word:
samplers which generate other samplers, probabilistic programs sampling the space of
probabilistic programs. 

In particular, one should mention 
the recent work on learning probabilistic programs by Yura Perov and Frank Wood~\cite{YPerovWood}
and recent advances in compositional concept learning obtained by Brenden Lake~\cite{BLake} (see Section 3 of~\cite{MBukatinMatthews}  for the brief
overview of these two sets of results).

We'll return to these two possible ways to understand the notion of ``higher-order programming"
for stream-based program architectures in Section~\ref{higher}.

\subsection{Negation}

The availability of negation is not strictly necessary for the material of this paper. Nevertheless, one often finds that the presence of negation
(that is, the ability to have negative coefficients in linear combinations) is very useful computationally.
For animations, one simply sets zero at the appropriate gray level and uses the standard color inversion operation as
negation (we are using this operation in the Fluid demo examples). 

For probabilistic sampling, one needs to use two sampling channels, positive and negative, in order to implement negation.
The formalism allowing negative probabilities is convenient for this situation. The ability to take negation is often
closely linked with the ability to handle partial and graded contradictions (see the material in~\cite{MBukatinMatthews} and
references therein for more details).

\section{Bipartite Graphs and First-order Dataflow}\label{first}

For a static dataflow graph, and assuming no need to manage subgraphs and make their copies, we found it
convenient to use a bipartite dataflow graph model: two main types of nodes are data steams (in this case,
image streams) and transforms. Directed links can only go from data streams to transforms (the case of data stream being a source for a transform),
or from transforms to data streams (target nodes for transforms). Links from data streams to data streams,
or from transforms to transforms are not allowed. 

In a typical situation, many data streams can have links into the same transform, but a transform only generates
one data stream. Exceptions from this typical situation exist, nevertheless this points to a certain kind of
duality: in a typical situation a data stream has one incoming graph edge and many outgoing graph edges,
while a transform has many incoming graph edges and one outgoing graph edge.

Invoking neural analogies, a data stream is similar to an axon, and a transform is similar to a body of a neuron together with its dendrites.

One often wants to modulate the transforms with the controls. One can think about controls
as special kinds of data streams (in a typical situation, the values stay constant until changed by a mouse
click). Unlike images, the amount of data in a control is usually small.

In this study, most dataflow computations are synchronous and are coordinated by global clock.
The stream of values of global clock is a special type of control (it is not clickable, but just
advances at a constant rate).

\subsection{Fluid Implementation}\label{first_fluid}

Implementation of this architecture in Fluid is rather straightforward: under 300 lines of code in
Processing~\cite{Processing} + about 70 lines of code for each of
the example dataflow graphs (see {\tt may\_9\_15\_experiment} subdirectory of the Fluid project).

The object oriented design of this experiment is as follows. The program $P$ is an object of {\tt MasterConfig} class,
which is built during the initial setup.
The work cycle of the program at every clock tick is as follows. First, the system draws
all images registered with $P$ as outputs. Then, the system applies all
transforms registered with $P$ as transforms. Finally, the system shifts the target instances of images
into the source  position for all data streams registered as (dynamic) data with $P$. In this fashion,
a new frame is produced for all dynamic data streams on each work cycle.

Images/data streams are of two types. {\tt DataRectangle} objects contain constant images, which can
be used as sources for transforms (and visualized if desired). In principle, nothing prevents us
from adding input movies, but we have not done so for this experiment. The objects of class {\tt  DataRectangleDynamic}
inherit from {\tt DataRectangle}, and also provide the target instances of images for the transforms to write to, and
the ability to shift the target and source images on each step of the work cycle.

Transforms all inherit from the abstract class {\tt Transform} with the single method {\tt apply}.
Three types of transforms turn out to be sufficient to provide rich behavior of the demos.
The negation transform reads the image on top of the source stream and produces its color inversion on the output.
The sum-of-2 transform reads images which are on top of its two source streams and
produces on the output their convex sum, $(1-\alpha)*A+\alpha*B$ where $0 \leq \alpha \leq 1$. The value of $\alpha$ is determined
by the associated instance of {\tt NumericControl}, which tends to be located to the right
of the drawing field of the image stream generated by the sum-of-2 transform in question.

{\tt CustomWaveTransform} reads the image on the top of the source stream and
produces an image on the output which can be thought of as a ``reflection in a snapshot of
a synthetic water wave". The parameters of this synthetic wave depend on the
implicit global clock control {\tt frameCount} built into Processing, and
also on the values stored in the associated {\tt CustomClickControl}.
Therefore, a wave transform takes an input stream of images and produces
a stream of the reflections of the input stream in the moving waves
(no physical realism is attempted).  The {\tt CustomClickControl}
associated with the wave simply uses the drawing field of the image stream generated
by the wave transform in question in order to catch clicks. 

Upon click, the wave is effectively restarted in the
position of the click as the new center, and the {\tt frame\_count\_base} of the control
is set to the current {\tt frameCount} (the wave dynamics depends only on {\tt frameCount - frame\_count\_base}).

All controls inherit from {\tt ClickControl} class and need to be registered with
the program as controls in order to work.

Two examples of a first-order dataflow program are included in the subdirectory\linebreak {\tt may\_9\_15\_experiment},
one is based on a directed acyclic graph, and one contains a loop. A short video
demonstrating work with the first of these examples is posted on {\tt youtube.com} (see the main page of the Fluid repository~\cite{Fluid}
for details).

\section{Almost Continuous Transformations and Streams of Dataflow Graphs}\label{continuous}

In this study, we support ``almost continuous evolution" of dataflow graphs while they are
running as programs. {\em Almost continuous evolution} of dataflow graphs is understood
as continuous evolution of their coefficients punctuated by {\em benign discontinuities}
at discrete moments of time.

One type of benign discontinuity is creation of new dataflow circuits which don't have
outgoing links to the currently existing and running dataflow circuits and thus don't affect
the existing behavior (unless the outside environment provides feedback based
on the output of the program). We find it convenient to implement this type of benign discontinuity
not by building new circuits vertex by vertex, but by copying the intact dataflow graphs
and subgraphs (see {\em Limited deep copy} in Section~\ref{limited_deep_copy} below).

It turns out the bipartite graph approach of the previous section makes it difficult
to correctly identify and copy subgraphs. Hence we adopt an entirely different
architecture, which is grouped around the notion of
{\em dataflow vertex} (also known as {\em target node}), with necessary transforms, controls, references
to sources, and associated datastreams all being subordinate to the target nodes
they belong to.

Two other kinds of benign discontinuity mentioned in the Introduction are combined
into an {\em S-insert} operation (see Section~\ref{special_insert} below for details
and discussion).

All benign discontinuities described here are reversible, and it seems likely that
one can almost continuously deform any dataflow graph into any other via
these benign discontinuities, their reversals, and continuous changes of
coefficients, as long as one type of data streams and a fixed set of built-in
transform operations is used.

\subsection{Dataflow Programs, Dataflow Graphs, and Dataflow Vertices}

In this study, a {\em dataflow program} $P$ is a finite collection of dataflow graphs, dataflow vertices, 
and data associated with dataflow vertices. This collection varies with time and is organized as folllows.

A {\em dataflow graph} $G$ in $P$ consists of a finite, possibly empty collection $\mathbf{V}_G$ of dataflow vertices in $P$ called {\em immediate target nodes} of
$G$ and a finite, possibly empty collection $\mathbf{S}_G$ of dataflow graphs in $P$ called {\em immediate subgraphs} of $G$.

For each $V \in \mathbf{V}_G$ graph $G$ is called a {\em parent graph} of $V$. For each $S \in \mathbf{S}_G$ graph $G$ is called a {\em parent graph} of $S$. 
In this study we require that for every vertex $V$ in a given program $P$ there is one and only one parent graph in $P$, and that for every
dataflow graph $G$ in a given program $P$ there is no more than one parent graph in $P$.  A dataflow graph $G$ in $P$ which does not
have a parent graph in $P$ is called a {\em top-level graph} of $P$.

Typically, exactly one of the top-level graphs of $P$ is called the {\em main graph} of $P$ and is executed, while it varies in time.

A {\em dataflow vertex} $V$ in $P$ consists of a finite, possibly empty collection of references to other dataflow vertices in $P$ (those references
are called {\em sources} of $V$), a structure representing {\em data associated with} $V$, and a reference to its parent graph.

The set of all vertices of a dataflow graph can be obtained by a recursively defined flattening operation,
$F(G) = V_G \cup \{F(S)|S \in S_G\}$. The set of all vertices of a program is $F(P) = \bigcup \{F(G)|G$ is a top-level graph of $P\}$.

In the software implementation of this study, dataflow programs, dataflow graphs, and dataflow vertices are represented by
classes {\tt ProgramEditor}, {\tt DataFlowGraph}, and {\tt DataFlowVertex} respectively.

\subsection{Data Associated with Dataflow Vertices}

A structure $D$ called {\em data associated with a dataflow vertex} in $P$ consists of the reference to the unique dataflow vertex $V$ containing $D$,
and the data itself. In the software implementation of this study, data associated with a dataflow vertex are represented by subclasses of class {\tt VertexData}.

In this study, the most important type of data associated with a dataflow vertex is a stream of images, which comes in several varieties. Class {\tt VertexDataImage}
and its subclass {\tt VertexDataImageDynamic} correspond to classes {\tt  DataRectangle} and {\tt  DataRectangleDynamic} from Section~\ref{first_fluid} respectively.
The three types of transforms described in Section~\ref{first_fluid} and their associated controls are now subclasses of {\tt  DataRectangleDynamic}.

In Section~\ref{higher}, we also consider streams of dataflow graphs as data associated with a dataflow vertex.

\subsection{Limited Deep Copy}\label{limited_deep_copy}

{\em Limited deep copy} of a dataflow graph $G$ in $P$, where $G$ can be a top-level graph or a subgraph, is a dataflow graph $G'$ in $P$ constructed 
by the following algorithm in the current implementation. $G'$ can be added to $P$ as a top-level graph or as a subgraph to an existing dataflow graph in $P$.

Step 1. $G'$ is created as a recursive copy of $G$, by recursive traversal of $G$. Structures for data associated with dataflow vertices of $F(G)$ are copied, because
the corresponding data streams in $G$ and $G'$ might differ, so they need to be able to unfold in different memory spaces.
For each dataflow vertex $V$ in $F(G)$, a forward reference to the correspondent $V'$ in $F(G')$ is created.
The source references of $V$ at this stage are copied intact into $V'$. 

Step 2. Source references of dataflow vertices of $G'$ are updated, by recursive traversal of $G'$. If a vertex $V'$ in $F(G')$ has a source
link to vertex $W$, and if vertex $W$ has
a forward reference to vertex $W'$, then this source link is set to point to $W'$.

Step 3. The clean-up step.  Forward references set in Step 1 are reset to null values, by recursive traversal of $G$.

The result is that the references to the sources external to $G$ are preserved in the newly created copies, the internal structure of $G$ including dataflow vertices, 
references to sources within $F(G)$, and the data associated with the dataflow vertices in $F(G)$ are subject to the usual ``deep copy"
procedures, and no outgoing external links from $F(G')$ are created (in the current implementation this is facilitated by the fact that all explicit links
are from target nodes to their sources; if a different implementation were to require explicit outgoing links, one would then need to take
special care to only include internal outgoing links in the ``deep copy" and to omit the external outgoing links).

\subsection{S-insert}\label{special_insert}

We have described the following two {\em benign discontinuities} in the Introduction.

 First,
one takes a datastream node $A$ and replaces it with nodes $B$ and $C$ connected by an identity transformation. Node $B$ inherits from node $A$ its
incoming links and their transformation. Node $C$ inherits from node $A$ its outgoing links. Second, one replaces $C=Id(B)$, with
$C= (1-\alpha)\cdot B + \alpha \cdot D$ where $\alpha = 0$, so a zero weight link from some node $D$ to $C$ is added. 

{\em S-insert} (soft insert/special insert) combines these two steps without going through the intermediate state of explicitly having $C=Id(B)$ configuration. The two main
parameters of this operation are dataflow vertices {\tt target\_vertex} and {\tt side\_vertex}. First a dataflow vertex {\tt new\_vertex} is created
and references to sources and to the parent graph are copied from {\tt target\_vertex} to  {\tt new\_vertex}. Then vertex data associated with
{\tt target\_vertex} are {\em moved} to {\tt new\_vertex}. Then {\tt target\_vertex} obtains two new sources,  
{\tt new\_vertex} and {\tt side\_vertex}, instead of its old sources. Then new vertex data based on sum-of-2 transform is created for
{\tt target\_vertex} and the coefficients of this transform are set in such a way that 1 corresponds to {\tt new\_vertex}
and 0 corresponds to {\tt side\_vertex}.

If the time step were infinitely small (a system similar to differential equations), 
this transformation would not have any immediate effect on the dynamics, which is
why we call it {\em benign discontinuity}. However, one should keep in mind that the replacement of
node $A$ by $C=Id(B)$ actually introduces time delay of one clock tick here in our
typical situation of discrete time. Normally one would expect the effect of this
extra time delay not to be too noticeable, but certainly there are some situations
when precise synchronization might matter, and then the operation is no longer
innocent.

The main effect of the {\em S-insert} operation is, however, that it enriches the space of
possibilities to continuously change the system, by allowing us to continuously
vary $\alpha$. 

\subsection{An Experimental Setup in Fluid}

Implementation of this architecture in Fluid is around 700 lines of code in Processing
 (see {\tt jun\_21\_15\_experiment} subdirectory of the Fluid project).

The setup of this experiment starts with the empty main graph (program) and
two small {\em template} top-level graphs, each consisting of two nodes, a constant image node
being a source for a transform node (in one case, the negation transform node,
in another case, the wave transform node).

\subsection{Program Execution in Fluid}\label{continuous_fluid}

The work cycle at every clock tick consists of two steps. One is execution of the work cycle
of the main graph (which is well defined even when the main graph is empty and is similar to the
work cycle described in Section~\ref{first_fluid}), and another is
an optional transformation of a program by benign discontinuities (in this experiment,
this transformation is programmed in the {\tt tweak\_optionally} method of the {\tt ProgramEditor}
class, eventually one should be able to also associate an edit control with the graph node containing
the program itself (see Section~\ref{higher})).

In this particular experiment, a limited deep copy of a wave transform {\em template} is added first,
then while the program is running, a limited deep copy of the negation
transform {\em template} is added at a later point, and then while the program is running, the S-insert is used to link
the output of the copy of the wave transform template to the input of the copy of negation transform template.
Section~\ref{higher} describes a follow-up experiment visualizing this program evolution (see Fig. 1).

While the program is running, one can also change wave parameters and linear coefficients by clicking on
controls which are associated with the drawing fields of the respective
target nodes in this experiment. Those changes constitute more abrupt discontinuities, 
so strictly speaking when those happen, it's not almost continuous anymore. 
But they only affect the parameters of the transformations, 
not the structure of the program, so we still tend to consider them to be within this programming paradigm overall.

\section{Higher-order Dataflow Programming}\label{higher}

There are two major classes of approaches to higher-order stream-based programming.
One approach takes standard higher-order functional programming as a starting point,
and focuses on integrating stream-based programming into the standard higher-order paradigm.
The second approach takes the notion of streams of programs as
its starting point and develops from there. The third approach which must be
mentioned is an approach based on multidimensional streams~\cite{WWadge}.

This second approach is the one we are
pursuing in this paper. We have already mentioned the dichotomy between the first two approaches in Section~\ref{sampling_the_samplers}
for the case of probabilistic programming. The body of research on higher-order dataflow
programming based on streams of functions is modest. An early work which should be mentioned
in connection with this is the preprint~\cite{SMatthews} by the second author. There are recent related
papers exploring various aspects of this approach, for example~\cite{NKrishnaswami}.

The previous section introduces streams of dataflow graphs which is the first step towards adopting this approach to higher-order dataflow programming
for Fluid.

The next step is to include such streams of dataflow graphs as datastreams contained in the nodes of other dataflow graphs (self-reference is, of course,
possible and is a natural starting point as we'll see in a moment). 

Pragmatically speaking, one needs to add the capabilities for imaging a dynamically changing dataflow graph, and it is desirable to
enable the use of such an image as an edit control capable of editing the underlying dataflow program while it is running
 (just like we use drawing fields of ordinary moving images as controls altering their dynamics in the examples above). It is
easier to achieve those goals if one includes a node containing a reference to the main graph of the program into the main graph
itself.

Our most recent experiment represents a partial step in this direction. A  node containing a reference to the main graph of the program
is included  into the main graph itself, but this setup is only used to provide visualization of a dynamically evolving dataflow
graph, there is no associated edit control yet  (see {\tt jun\_28\_15\_experiment} subdirectory of the Fluid project).

One starts with the main graph containing only the node containing a reference to the main graph itself, and then
gradually adds functionality as in Section~\ref{continuous_fluid}.
A short video demonstrating gradual evolution of this data flow program while it is running and being used in an interactive
fashion is posted on {\tt youtube.com} (see the main page of the Fluid repository~\cite{Fluid} for the link
to the video and see the legend included with the video for details; see Fig. 1 for a screenshot).

\begin{figure}\label{bukatin_figure}

\centering
\includegraphics[scale=0.225]{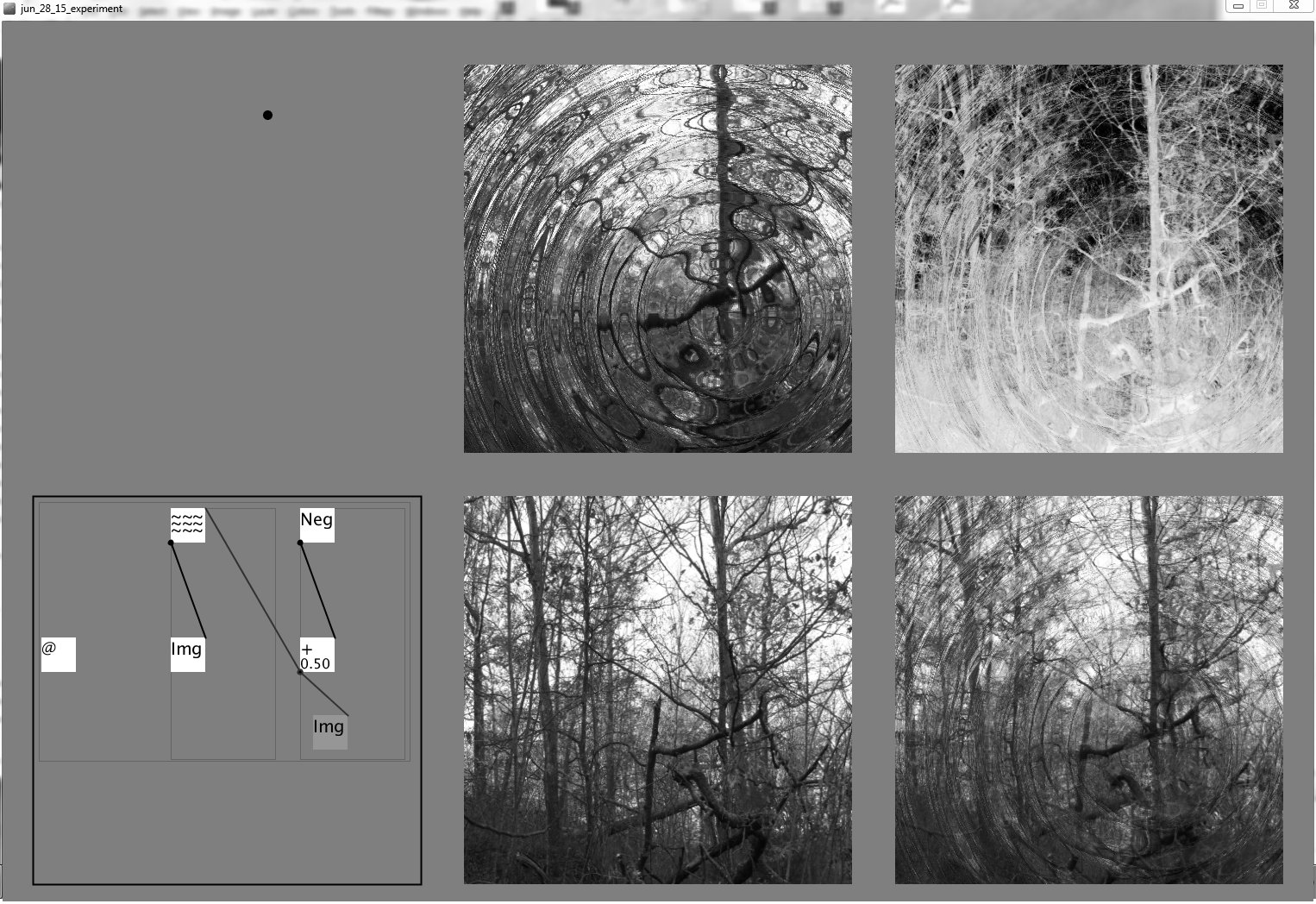}

\caption{Screenshot from {\tt jun\_28\_15\_experiment} }

\end{figure}

\subsection{Further Work in Higher-order Dataflow Programming}

This is a very modest start on the road towards real-life higher-order dataflow programming and program evolution
involving streams of almost continuously changing programs. One would like the node containing a dataflow graph
to depend on other nodes as sources.

Then in a higher-order setting one could, for example, modulate the evolution and behavior of a stream of dataflow graphs by a moving image, etc.

One circumstance to keep in mind is that in order to evolve a program while it is running, the changing program must inherit state from the earlier
moments in time, just like it is done in Section~\ref{continuous}. This seems to imply that the node containing a dataflow graph must be one of
its own sources.

\section{Conclusion}

Most of the formalism we described is applicable to any kind of streams. However, the ability to evolve streams of programs via almost continuous
transformations requires the ability to take convex linear combinations of two streams. In the absence of such an ability, one might need
to rely on discontinuous transformations, such as abrupt switching of a link from one source to another.

The ability to almost continuously evolve dataflow programs while they are running makes it possible
to sample almost continuous trajectories in the space of dataflow programs. This should allow us to try new evolutionary and
probabilistic schemas for
program learning.

Stream based architecture is conductive to various modalities such as audio streams, streams of text snippets, etc., and also for
streams of samples from probability distributions of arbitrarily structured objects. Many of them, in particular audio steams and
streams of probabilistic samples, admit convex linear combinations.

The ``code sensory modality" (see Section 3.1 in~\cite{EYudkowsky})
is the most fundamental one, since everything in the digital world is made from code. Moving from discrete
representations of code to streams of code, and, in particular, to almost continuous streams of code is likely to be fruitful
in this context.

{\bf Acknowledgments.} We would like to thank Ralph Kopperman, Lena Nekludova, and Josh Tenenbaum for helpful discussions.

\end{document}